\documentclass[lettersize, journal]{IEEEtran}
\usepackage[withpage]{acronym} 
\usepackage{xcolor,soul}
\sethlcolor{yellow}

\usepackage{amsmath,amsfonts}
\usepackage{array, tabularx}
\usepackage{algorithm}
\usepackage{subfigure}
\usepackage{caption}
\usepackage{textcomp}
\usepackage{stfloats}
\usepackage{url}
\usepackage{verbatim}
\usepackage{graphicx}
\usepackage{cite}
\usepackage{hyperref}
\usepackage{float}
\usepackage{academicons}
\usepackage{hyperref}
\newcommand{\orcidauthorA}{\href{https://orcid.org/0000-0001-5792-0842}{\includegraphics[scale=0.05]{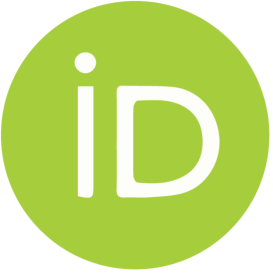}}}
\newcommand{\orcidauthorB}{\href{https://orcid.org/0000-0003-4466-0089}{\includegraphics[scale=0.05]{Figures/orcid_16x16.eps}}}
\newcommand{\orcidauthorC}{\href{https://orcid.org/0000-0003-3836-1373}{\includegraphics[scale=0.05]{Figures/orcid_16x16.eps}}}
\newcommand{\orcidauthorD}{\href{https://orcid.org/0000-0002-1478-2272}{\includegraphics[scale=0.05]{Figures/orcid_16x16.eps}}}

\begin{document}

\title{Energy-Efficient and Reliable Data Collection in Receiver-Initiated Wake-up Radio Enabled IoT Networks}

\author{Syed Luqman Shah$^{\orcidauthorA{}}$, Ziaul Haq Abbas$^{\orcidauthorB{}}$, Ghulam Abbas$^{\orcidauthorC{}}$, \textit{Senior Member, IEEE}, and Nurul Huda Mahmood$^{\orcidauthorD{}}$
\thanks{S. L. Shah and N. H. Mahmood are with the 6G Flagship, Centre for Wireless Communications (CWC), University of Oulu, 90014 Oulu, Finland. E-mail: syed.luqman@oulu.fi, nurulhuda.mehmood@oulu.fi\\
Z. H. Abbas is with the Faculty of Electrical Engineering, GIK Institute of Engineering Sciences and Technology, Topi 23640, Pakistan. G. Abbas is with the Telecommunication and Networking (TeleCoN) Research Center, GIK Institute of Engineering Sciences and Technology, Topi 23640, Pakistan. E-mail: ziaul.h.abbas@giki.edu.pk, abbasg@giki.edu.pk}
\thanks{Manuscript received May 15, 2025; revised July --, 2025.}}

\markboth{Journal of IEEE Class Files,~Vol.~--, No.~--, May~2025}%
{Shell \MakeLowercase{\textit{et al.}}: A Sample Article Using IEEEtran.cls for IEEE Journals}


\maketitle

\begin{abstract}
In unmanned aerial vehicle (UAV)-assisted wake-up radio (WuR)-enabled internet of things (IoT) networks, UAVs can instantly activate the main radios (MRs) of the sensor nodes (SNs) with a wake-up call (WuC) for efficient data collection in mission-driven data collection scenarios. However, the spontaneous response of numerous SNs to the UAV's WuC can lead to significant packet loss and collisions, as WuR does not exhibit its superiority for high-traffic loads. To address this challenge, we propose an innovative receiver-initiated WuR UAV-assisted clustering (RI-WuR-UAC) medium access control (MAC) protocol to achieve low latency and high reliability in ultra-low power consumption applications. We model the proposed protocol using the $M/G/1/2$ queuing framework and derive expressions for key performance metrics, i.e., channel busyness probability, probability of successful clustering, average SN energy consumption, and average transmission delay. The RI-WuR-UAC protocol employs three distinct data flow models, tailored to different network traffic conditions, which perform three MAC mechanisms: channel assessment (CCA) clustering for light traffic loads, backoff plus CCA clustering for dense and heavy traffic, and adaptive clustering for variable traffic loads. Simulation results demonstrate that the RI-WuR-UAC protocol significantly outperforms the benchmark sub-carrier modulation clustering protocol. By varying the network load, we capture the trade-offs among the performance metrics, showcasing the superior efficiency and reliability of the RI-WuR-UAC protocol.
\end{abstract}


\section{Introduction}
\label{Sec: Introduction}
Internet of things (IoT) networks are often deployed in hazardous, remote, inaccessible, or dynamic environments for smart monitoring, surveillance, and disaster response. In such cases, the battery-operated sensor nodes (SNs) that comprise these networks cannot be easily recharged or replaced \cite{IoTJ_Drone_Based_non_cellular}. To address this challenge, the integration of unmanned aerial vehicles (UAVs) into IoT networks, along with the development of efficient medium access control (MAC) protocols, has gained significant attention. UAVs are particularly advantageous due to their ability to rapidly deploy, establish line-of-sight (LoS) communication links, and access hard-to-reach areas \cite{IoT_Magzine}. UAVs can adjust their altitude to reduce data transmission distances, leading to decreased radio energy consumption and enabling real-time or near-real-time data collection, which also supports advanced learning algorithms to make timely decisions for many autonomous systems \cite{hu2023timely, real_time_AI_UAV_enable}.

On the other hand, MAC protocols manage the radio communication units (RCUs) of the SNs, which is the most power-intensive component among the sensing and micro-controller units (MCUs) \cite{WuR_IoTJ, ghose2019protocol, Luqman_EEUCH}. MAC protocols also enhance network reliability by preventing collisions and retransmissions of data frames and acknowledgments (ACKs), which can otherwise degrade network performance by increasing latency and the active time of the RCU, thereby reducing energy efficiency \cite{ghose2019protocol}. To address these challenges and minimize the active time of the RCU, two main types of MAC protocols have been proposed in the literature: (a) duty cycle (DC) and (b) wake-up radio (WuR) MAC protocols.

DC-MAC protocols were primarily developed to reduce energy consumption in SNs by alternating the RCU between active and deep sleep modes based on predefined time intervals, regardless of data communication demand \cite{zhang2017does, oller2015has}. However, DC-MAC protocols can lead to idle listening and overhearing during active modes, resulting in unnecessary energy consumption. Idle listening occurs when the RCU is active but not engaged in data communication, while overhearing occurs when the RCU listens to data frames not intended for the SN. In both cases, the RCUs waste energy.
To mitigate these issues, WuR MAC protocols have been developed for ultra-low power IoT applications \cite{magno2016design}, where the RCU of an SN is equipped with two radios: a main radio (MR) and a wake-up receiver (WuRx) \cite{WuR_IoTJ, zhang2017does}. In active mode, the MR is turned on which is used for transmitting and receiving data frames, wake-up calls (WuCs), and ACKs. During deep sleep mode, the MR is turned off to conserve energy, while the WuRx is always on to detect incoming WuCs \cite{WuR_Operation_IEEE_Standard}.\\
WuR-enabled SNs primarily remain in sleep mode, keeping their MR off until the MCU triggers it after detecting an authentic WuC with the WuRx, thus conserving $1000$ times more energy, as demonstrated in \cite{oller2015has, Spenza}. After successful data exchange, the SN reverts to deep sleep mode. The WuR approach operates on an on-demand basis, activating the MR only when an authentic WuC is received, thus completely eliminating idle listening and significantly reducing overhearing. To further reduce overhearing early sleeping and early data transmission is proposed in \cite{ghose2019enabling}, which involves validating the WuC frame address bit-by-bit before triggering the MR of the intended SN.

In IoT networks, star and tree topologies are typically used for physical environment monitoring \cite{Water_monitoring, ghose2018mac, hsu2021receiver}, with a centralized node called the cluster head (CH) and leaf nodes called cluster members (CMs) \cite{Luqman_EEUCH, ghose2018mac, heinzelman2000energy}. In WuR-IoT networks, communication can be initiated by either the transmitter (e.g., the CMs) or the receiver SN (e.g., the CH) via a WuC, enabling WuR MAC protocols to operate in either transmitter-initiated (TI) or receiver-initiated (RI) modes \cite{ghose2019protocol, ghose2018mac, WuR_Operation_IEEE_Standard}. In TI-WuR mode, the transmitter SN (i.e., the CM) initiates communication by sending a WuC to the receiver SN (i.e., the CH). Upon receiving the WuC, the receiver CH SN's MCU activates its MR to receive data frames. Once data transmission is complete, the MR is turned off, and the CH returns to sleep mode. TI-WuR mode is particularly suited for event-triggered data reporting scenarios, such as industrial automation, environmental monitoring, smart agriculture, and structural health monitoring, where SNs remain in deep sleep mode until a specific event prompts them to send WuCs to wake up the intended receiver CH.\\
Conversely, in RI-WuR mode, the receiver CH SN initiates communication by sending a WuC to the transmitter CM SN. The transmitter CM SN's WuRx detects the WuC and activates its MR to send data frames to the intended receiver CH. RI-WuR mode is ideal for mission-driven data collection scenarios, such as smart agriculture, urban planning and smart cities, surveillance, and disaster response, where the receiver collects data from WuR-enabled deployed SNs to gather data on a regular or on-demand basis, rather than waiting for specific events to occur. The collected data is then transmitted to a central monitoring or control system, where the data are analyzed and used for decision-making, resource optimization, or regulatory compliance \cite{ghose2019protocol}.

Although, WuR technology is dessigned for ultra-low-power IoT applications, its effectiveness diminishes in high-traffic environments \cite{ghose2018mac, magno2016design, zhang2017does, oller2015has}. In mission-driven data collection scenarios, where a UAV triggers the MRs of deployed SNs with a WuC for data collection in remote or hazardous areas, the concurrent response of multiple SNs to the UAV's WuC can overwhelm the network \cite{hsu2020achieving}. This congestion leads to significant packet loss and collisions, as WuR was not initially designed for high-traffic loads \cite{ghose2019protocol, ghose2019enabling}. To address this gap, we introduce the receiver-initiated wake-up radio UAV-assisted clustering (RI-WuR-UAC) protocol, a highly energy-efficient and reliable handshake mechanism explicitly designed for UAV-assisted clustering in such high-traffic environments.

\subsection{Contributions}
The proposed RI-WuR-UAC protocol operates on an on-demand basis, allowing UAVs to act as CHs that form independent, non-overlapping clusters of SNs within the field of interest (FoI). It supports different network traffic conditions by incorporating three distinct data flow models tailored for light, heavy, and variable network loads. RI-WuR-UAC operates in independent rounds, each beginning with the arrival of the UAV at the FoI. Each round consists of two phases: the setup phase, and the steady-state phase.

During the setup phase, the UAV broadcasts a WuC to WuR-enabled SNs, that are initially in deep sleep mode. Upon receiving the WuC, the SNs' WuRx activate their MR via the MCU, allowing them to respond with joining request (JReq) frames. The UAV then forms a cluster of those SNs whose JReq frames are received, marking them as its CMs, and assigns time division multiple access (TDMA) slots to each of its CM to enable collision-free data collection during the steady-state phase. In the steady-state phase, the UAV collects data frames from its CMs in their assigned TDMA slots. After successful data collection, the UAV transmits an ACK frame to the CM SNs. The proposed RI-WuR-UAC protocol makes the following key contributions:
\begin{itemize}
\item The RI-WuR-UAC protocol enables UAVs to act as CHs that form independent, non-overlapping clusters within an FoI. This approach minimizes energy consumption of the SN and data aggregation/computation at CH during the setup phase.
\item To handle various network traffic conditions and enhance the protocol's applicability in diverse environments, we introduce innovative data flow models that incorporate three distinct MAC mechanisms, each tailored for different network loads:
\begin{enumerate}
    \item Clear channel assessment (CCA) clustering for light network loads.
    \item IEEE 802.15.4 non-beacon mode unslotted carrier sense multiple access with collision avoidance (CSMA-CA) clustering for congested network loads.
    \item Adaptive clustering for variable network loads.
\end{enumerate} 
\item We develop mathematical framework for $M/G/1/2$ queue to derive key performance metrics, i.e., channel busyness probability, the probability of successful cluster formation, the average transmission delay, and the average energy consumption, for the RI-WuR-UAC protocol. These expressions are evaluated for the proposed protocol and simulated across the three data flow models to assess performance under different network loads and varying cluster sizes.
\end{itemize}

 \begin{table*}[h]
	\caption{Comparision of proposed RI-WuR-UAC and existing RI-WuR protocols.}
	\begin{center}
		\begin{tabular}{|>{\centering\arraybackslash}m{6.9cm}|>{\centering\arraybackslash}m{10cm}|}
			\hline
			RI-WuR-UAC [Proposed]	&  Protocols in~\cite{hsu2021receiver,hsu2020achieving}\\
			\hline \hline
			SNs are not classified into groups.	& SNs are divided into four groups.\\
			\hline
			Data frames are collected for every SN that a single WuC covers.	& From all of the SNs covered by a single WuC, a single data frame is collected from each SN in the specified group of SNs.\\
			\hline
			All of the data frames are gathered in a single round.	& 	One data frame is collected from each SN in specific group of SNs, while the remaining SNs from the other groups in the same cluster region must wait until the next round.\\
			\hline
			The UAV determines the number of TDMA slots based on the number of JReqs received. 	& The TDMA slot is the result of a hash function calculated by each SN. It is obvious for a collision to occur if the hash function's output for two or more SNs is the same; the likelihood of happening this increases with the number of SNs in a group.\\
			\hline
			More than one data frames can be collected from each SN in a signal TDMA slot because the number of the TDMA slots assigned to each SN is application dependent. In particular, the entire data from a single SN is gathered during a single round. & To accommodate SNs, groups are made. One frame per round per SN from a specific group is collected in a single TDMA slot. Due to the possibility of an SN having multiple frames, this method results in an increase in latency because these SNs have to wait for the next turn. Furthermore, SNs belonging to other groups must wait for the subsequent rounds.\\
			\hline
			Each SN sends just one JReq on behalf of the WuC to the particular UAV, which prevents clusters from overlapping. 	& The overlapping of clusters is eliminated using a complex partition algorithm.\\
			\hline
		\end{tabular}
		\label{Table:1}
	\end{center}
\end{table*}
 
\subsection{Paper Organization}
After providing a brief introduction in Sec. \ref{Sec: Introduction}, we review the existing state-of-the-art literature in Sec. \ref{Sec: Related Work}. Sec. \ref{Sec: System Model} describes the network model, the operation of the proposed protocol, and the data flow models. In Sec. \ref{Sec: Protocol Modeling}, we present the analytical model of the proposed protocol, followed by the derivation of key performance parameter expressions in Sec. \ref{Sec: Parameters for Performance Analysis}. Sec. \ref{Sec: Simulations} discusses the results obtained, and we conclude the work with potential future directions in Sec. \ref{Sec: Conclusions}.

\section{Related Work} \label{Sec: Related Work}
The development of the RI-WuR-UAC protocol builds upon the existing research on WuR protocols. Key studies include work on TI WuR \cite{ghose2018mac, ghose2019enabling, ghose2019protocol} and RI WuR \cite{Luqman_EEUCH, hsu2020achieving, hsu2021receiver}. Furthermore, benchmark protocols capable of operating in both TI and RI WuR modes are presented in \cite{SCM-WuR1, oller2015has}. These studies have paved a way for the development of the proposed RI-WuR-UAC.\

The sub-carrier modulation (SCM)-WuR, also known as correlator (Cor)-WuR~\cite{ghose2018mac, oller2015has, Spenza}, is a widely recognized benchmark for evaluating WuR MAC protocols~\cite{ghose2019protocol, ghose2018mac}. SCM-WuR supports both TI and RI modes of operation. In our work, SCM-WuR serves as the reference protocol for assessing the RI-WuR-UAC's performance in RI mode. SCM-WuR lacks MAC mechanisms, such as back-off (BO) or CCA, before transmitting any frame. RI-WuR-UAC introduces specific MAC strategies to address these shortcomings of the RI-mode. Prior work on SCM-WuR~\cite{SCM-WuR1, oller2015has} focuses on TI mode communication, where a WuR-enabled SNs initiate data transmission by sending a WuC to the intended receiver. Also, in \cite{ghose2018mac}, the Cor-WuR is used as a benchmark for evaluating their proposed protocols, i.e., CCA-WuR, CSMA-CA WuR, and ADP-WuR.

In event-triggered data reporting scenarios, multiple SNs may simultaneously detect and report anomalies, leading to potential collisions and heavy packet loss. To address this issue, the authors in~\cite{ghose2018mac} proposed three TI-WuR MAC protocols for WuR-enabled IoT networks: (a) CCA WuR, (b) unslotted CSMA-CA WuR in non-beacon mode, and (c) adaptive (ADP) WuR. These protocols are designed to minimize heavy packet loss in TI-mode of operation by implementing various collision avoidance (CA) mechanisms, which are notably absent in traditional WuR protocols. Their effectiveness was evaluated using the $M/G/1/2$ queue framework, demonstrating a reduction in collisions, successful cluster formation, and improved reliability during simultaneous data transmissions.

In contrast, mission-driven data collection scenarios, such as those found in~\cite{hsu2021receiver, hsu2020achieving}, involve collecting data in WuR-enabled IoT applications using UAVs. The authors proposed a hashing and partitioning approach to manage data collection, which involved dividing SNs within a UAV's cluster region into groups. However, this approach has notable limitations. Each UAV collects only one data packet from each SN in a specific group per round, forcing SNs in other groups to wait for subsequent rounds. TDMA slots are assigned based on a hash function embedded in the broadcasted WuC \cite{hsu2021receiver}. While this method assigns slots, it can also lead to collisions if two or more SNs generate the same hash output, with the likelihood of collisions increasing as the number of SNs in a group grows. This approach not only increases latency but also results in inefficient UAV energy use, as SNs with multiple packets must wait for subsequent rounds to transmit. Furthermore, a complex partition algorithm is used to eliminate overlapping clusters, further complicating the system. Our proposed RI-WuR-UAC protocol offers a significantly different methodology by eliminating the need for cluster-based data collection and complex partition algorithms, as detailed in our previous work~\cite{Luqman_EEUCH}. Table~\ref{Table:1} summarizes the key differences between the proposed RI-WuR-UAC protocol and the protocols described in~\cite{hsu2021receiver, hsu2020achieving}.

The RI-WuR-UAC builds upon our previous work \cite{Luqman_EEUCH}, where we introduced the energy efficient UAV-assisted clustering hierarchical (EEUCH) routing protocol. The EEUCH protocol was evaluated across eleven performance metrics, including the number of alive and dead SNs, total packets collected from the FoI, remaining network energy, average network energy reduction, network energy consumption, network lifetime, stability time, average throughput, control overhead, and the information coverage ratio of the UAV. Results demonstrated the EEUCH protocol's outstanding performance across these metrics. Compared to protocols like Low energy adaptive clustering hierarchy (LEACH) \cite{heinzelman2000energy}, LEACH-centralized (LEACH-C) \cite{LEACH_C}, stable energy protocol (SEP) \cite{SEP}, threshold sensitive energy effective sensor network (TEEN) \cite{TEEN}, improved energy-efficient LEACH (IEE-LEACH) \cite{IEE_LEACH}, and enhanced energy-efficient clustering approach for three-tier heterogeneous wireless sensor networks (EEECA-THWSN) \cite{THWSN}, the EEUCH protocol significantly extended the lifespan of SNs by 12.55, 12.13, 9.55, 6, 8.92, and 4.72 times, respectively. SNs using the EEUCH protocol maintained their energy for up to 1080 rounds, outperforming other protocols, where SNs exhausted their energy as early as rounds 86, 89, 113, 180, 121, and 229. The EEUCH protocol excelled in packet collection, gathering 11.74, 7.78, 9.05, 11.22, 5.10, and 5.06 times more packets than LEACH, LEACH-C, SEP, TEEN, IEE-LEACH, and EEECA-THWSN, respectively. Furthermore, it achieved better average throughput, transmitting 11.74, 7.78, 9.05, 11.22, 5.11, and 5.06 times more bits per round compared to these protocols. The control overhead generated by the EEUCH protocol was measured at 11\%. The EEUCH protocol demonstrates how independent, non-overlapping clusters can be formed and how a large volume of data packets can be collected effectively. However, it does not address critical MAC layer responsibilities, such as how SNs can access the shared wireless medium, avoid collisions during the setup phase, ensure successful cluster formation, or determine the time and energy required for cluster formation. The proposed RI-WuR-UAC protocol extends the MAC layer functionalities of the EEUCH protocol, addressing these unanswered questions. RI-WuR-UAC also incorporates three distinct data flow models designed to handle light, heavy, and variable network loads.

\section{System Model}
\label{Sec: System Model}
This section presents the network scenario and assumptions for the proposed RI-WuR-UAC protocol, which incorporates three data flow models designed to handle different network load conditions. It also explains the operation principles of the proposed protocol.

\subsection{Network Scenario and Assumptions}
This subsection offers a description of the network model and outlines the key assumptions made at the MAC layer for the performance evaluation of the RI-WuR-UAC protocol.\
 \begin{figure}[b]
	\centerline{\includegraphics[width=0.49\textwidth]{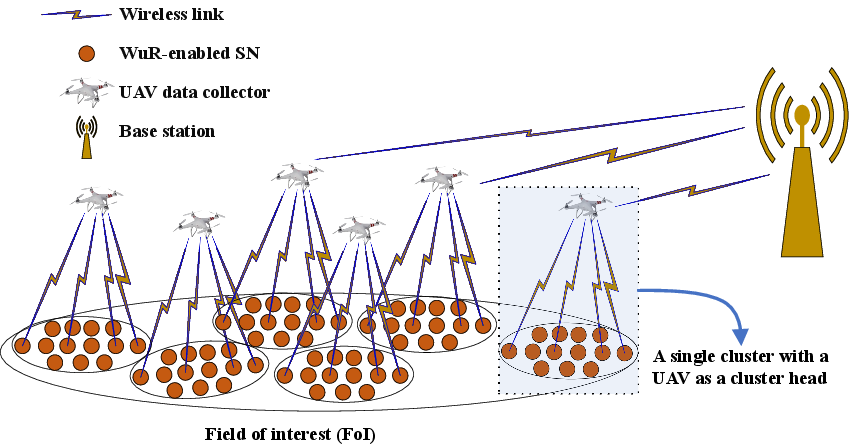}}
	\caption{The proposed system model highlighting the single cluster formation.}
	\label{Ch_4 System Model}
\end{figure}

\begin{figure}[t]
	\centerline{\includegraphics[width=0.49\textwidth]{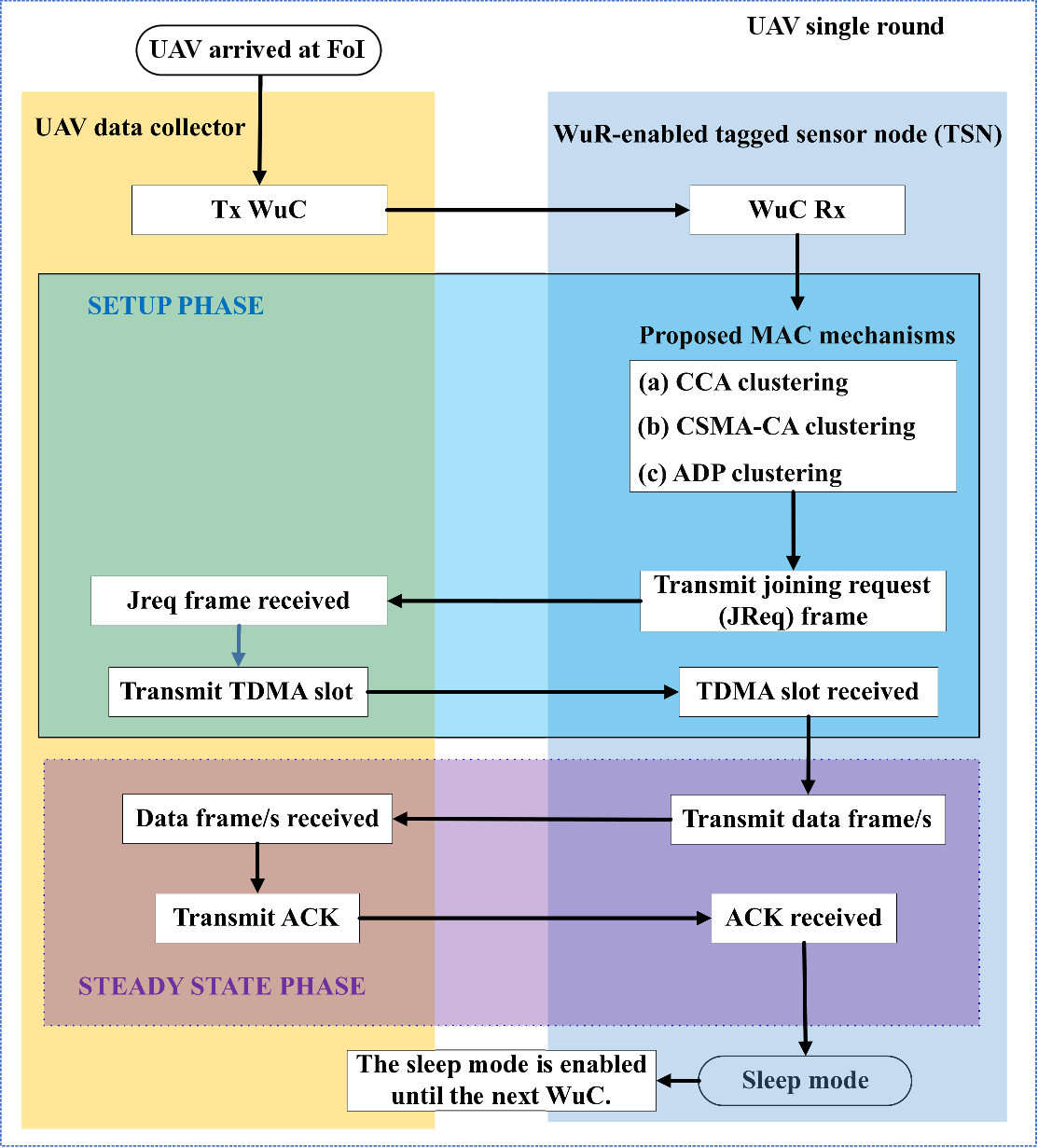}}
	\caption{Illustration of a single round of the RI-WUR-UAC protocol's working principle, highlighting the processes occurring at the UAV and WuR-enabled SNs in the right and left vertical big blocks, respectively. The setup phase and steady-state phase of the protocol are highlighted in the top and bottom horizontal big blocks, respectively.}
	\label{Ch4 Operation at each SN}
\end{figure}

Consider a single cluster containing $N$ WuR-enabled SNs, with a UAV acting as the CH responsible for data collection. This network is assumed to be deployed in a hazardous, remote, or dynamic environment for applications such as monitoring, surveillance, or disaster response, as highlighted in the rectangular frame in Fig. \ref{Ch_4 System Model}. Each SN is equipped with a WuRx and an MR, and has a finite queue capacity to efficiently manage and store data frames for transmission. Due to limited accessibility in such environments, UAVs are employed to collect data from these battery-operated SNs. Upon arriving at the FoI, the UAV broadcasts a WuCs to the WuR-enabled SNs. Upon receiving the WuC, the SNs respond by transmitting a JReq frame back to the UAV. The UAV collects a number of JReq frames from the SNs, marks those SNs as its CMs and assigns TDMA slots to each CM for reliable data transmission. The CMs send data frames to the UAV during their assigned TDMA slots, enabling the UAV to collect data reliably. This process is illustrated in Fig. \ref{Ch4 Operation at each SN}.

It is important to note that the FoI may consist of multiple clusters, depending on its physical size and the number of deployed SNs, as depicted in Fig. \ref{Ch_4 System Model}. In such cases, multiple UAVs will be required to cover the entire FoI, meaning that an SN may receive multiple WuCs from different UAVs. However, each SN responds only to the first WuC it receives by sending a JReq frame. This approach ensures that SNs do not join multiple clusters, thereby preventing cluster overlap. UAVs assign TDMA slots only to those SNs from which they have received JReq frames, making them CMs. As a result, clusters remain independent and non-overlapping, as detailed in the EEUCH routing protocol \cite{Luqman_EEUCH}. Given this independent and non-overlapping nature, this study specifically focuses on a single cluster to provide a detailed analysis of the proposed RI-WuR-UAC protocol.

For the sake of clarity, the following assumptions are made:
\begin{enumerate}
	\item The SNs are stationary, and each SN generates packets according to a Poisson process with a rate of $\lambda$.
	\item The number of data frames associated with each SN in each single round follows a uniform distribution ranging from 1 to 5.
	\item The channel under consideration is error-free and there are no hidden terminals within the cluster.
	\item The position of a UAV cannot be changed during the steady state phase of RI-WuR-UAC.
\end{enumerate}

\subsection{The RI-WuR-UAC MAC Protocol}
\begin{figure}[t]
	\centerline{\includegraphics[width=0.45\textwidth]{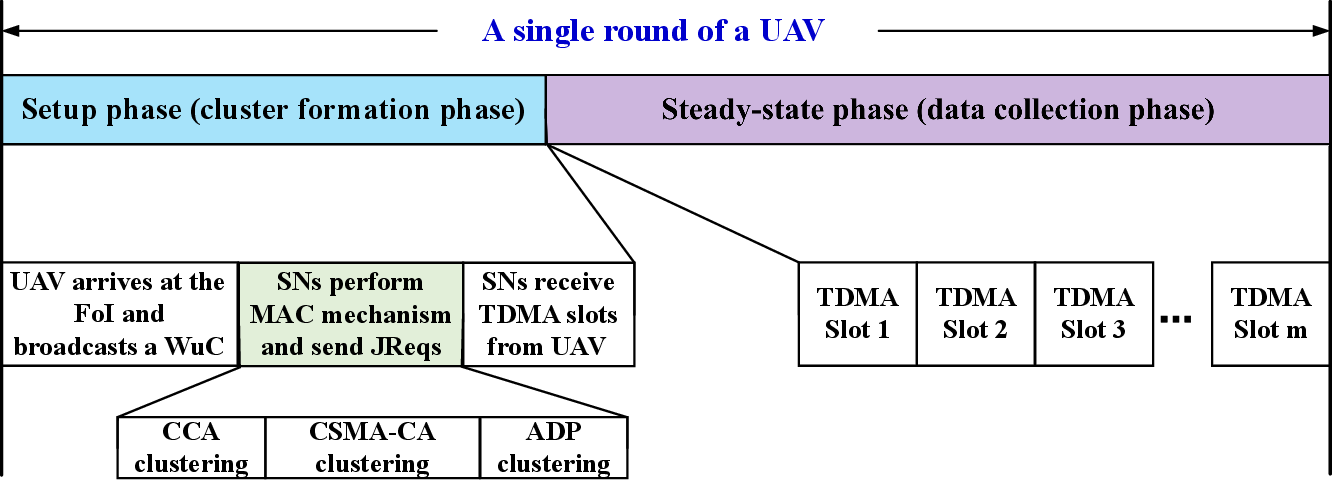}}
	\caption{Illustration of a single round of a UAV for the proposed protocol.}
	\label{Ch4 RI-WuR-UAC}
\end{figure}

This subsection provides an overview of the RI-WuR-UAC MAC protocol, focusing on a single UAV round. The round is divided into two main phases: the setup phase and the steady-state phase, as depicted in Fig.~\ref{Ch4 RI-WuR-UAC}\

\subsubsection{Setup Phase}
During the setup phase, the UAV arrives at its pre-defined hover position above the FoI and broadcasts a WuC signal. When the WuRxs of the SNs detect this WuC, the SNs activate their MRs after a brief mode-switching time (MST). This transition from sleep mode to active mode allows the SNs to respond to the UAV's WuC by sending JReq frames. In this case multiple SNs simultaneously send JReq frames in response to the WuC, so packets collisions can occur, leading to heavy packet loss. To mitigate this, the SNs compete asynchronously for access to the shared wireless channel, aiming to transmit their JReq frames to the UAV without collisions during this single-hop transmission.

The asynchronous operation mode allows SNs to independently contend for channel access, providing flexibility and adaptability within the network. To ensure reliable JReq frame transmission from SNs to the UAV, we propose three MAC mechanisms for the three data flow models tailored to different traffic loads: CCA clustering, IEEE 802.15.4 standard non-beacon mode unslotted CSMA-CA clustering, and ADP clustering. These mechanisms cater to clusters with light, heavy, and variable loads, respectively. The behavior of each MAC protocol is examined through a tagged SN (TSN), representing an SN within the cluster that competes with $N-1$ other SNs for channel access.\\

\begin{figure*}
    \centering
    \includegraphics[width=\textwidth]{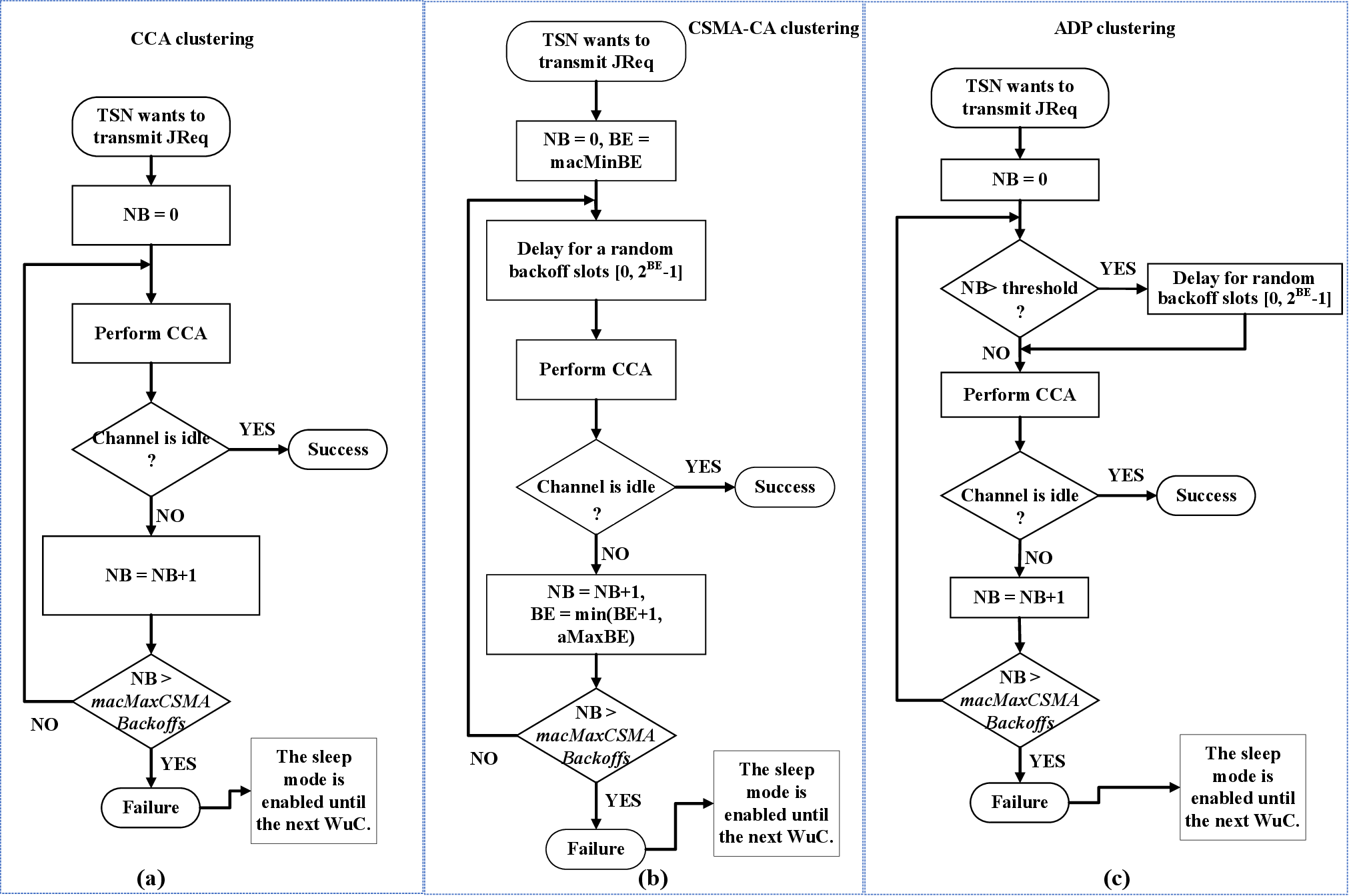}
    \caption{Working procedure of the (a) CCA clustering, (b) CSMA-CA clustering, and (c) ADP clustering for reliable transmission of JReq frame.}
    \label{fig4: Three MAC Mechanisms}
\end{figure*}

\noindent\textbf{CCA clustering:}\\
In CCA clustering, the TSN performs a CCA to determine whether the channel is available before transmitting its JReq frame. Unlike other MAC protocols, CCA clustering does not use a backoff (BO) algorithm. If the channel is idle during the CCA period, the TSN proceeds to send the JReq frame. If the channel is busy, the TSN continues performing the CCA process until it reaches the maximum retry limit. If the JReq frame transmission is unsuccessful within this limit, the TSN discards the JReq frame and notifies the higher layer of the network, which takes appropriate action. Consequently, the TSN drops out of the cluster formation process. If the TSN successfully gains channel access, it transmits the JReq frame and becomes a CM of the UAV. This process is illustrated in Fig.~\ref{fig4: Three MAC Mechanisms}(a).\\

\noindent\textbf{CSMA-CA Clustering:}\\
In CSMA-CA clustering, the TSN initiates a BO process before checking the channel's status to improve its chances of winning channel access. The BO process starts with a counter value randomly selected from the contention window (CW). As the counter decreases to zero, the TSN performs a CCA to check if the channel is idle. If the channel is idle, the TSN transmits the JReq frame. If the channel is busy, the TSN repeats the BO process followed by the CCA. If the JReq frame transmission is unsuccessful within the predefined retry limit of BO and CCA attempts, the TSN discards the JReq frame, notifies the higher layer, and drops out of the cluster formation process. Compared to CCA clustering, the BO process in CSMA-CA consumes less energy, i.e., $E_{BO} < E_{CCA}$ \cite{ghose2018mac, ghose2019protocol}. Under heavy traffic conditions, where the cluster is densely populated with SNs, CCA clustering requires multiple CCAs to access the channel, whereas CSMA-CA clustering may need only one or a few CCAs to win channel access. The CSMA-CA clustering operations are otherwise similar to those of CCA clustering, as shown in Fig.~\ref{fig4: Three MAC Mechanisms}(b).\\

\noindent\textbf{ADP clustering:}\\
The ADP clustering protocol ensures reliable JReq frame transmission under varying traffic load conditions by adaptively adjusting the MAC procedures of TSNs based on observed network traffic. In ADP clustering, the TSN monitors its JReq frame transmission attempts and compares them with a predefined threshold. Initially, the TSN employs the CCA clustering protocol but switches to CSMA-CA clustering if the attempt counter exceeds the threshold value ($t_h$). The $t_h$ can be dynamically configured based on past transmission statistics, such as JReq frame collisions, latency, and the number of attempts required for successful transmission. This dynamic threshold adaptation mechanism ensures the resilience of the MAC adjustment process. The working principle of the ADP clustering protocol is illustrated in Fig.~\ref{fig4: Three MAC Mechanisms}(c). If the number of BOs is below the predefined threshold, the ADP clustering protocol follows the CCA clustering process; otherwise, it follows the CSMA-CA clustering process.

Each SN from the cluster employs one of these MAC mechanisms based on the cluster size to which the TSN belongs. If the TSN is part of a light cluster, it uses CCA clustering. If it is part of a densely populated cluster, it uses CSMA-CA clustering. For clusters where the number of SNs varies, the TSN employs ADP clustering.\\

\subsubsection{Steady State Phase}

Once the UAV receives JReq frames from the SNs, it allocates TDMA slots to those SNs whose JReqs were successfully received. The number of TDMA slots assigned to each SN is based on the number of data frames associated with that specific SN, setting the stage for the steady-state phase. During the steady-state phase, SNs transmit their sensed data frames to the UAV within their allocated TDMA slots. Upon receiving these frames, the UAV sends an ACK to the SNs to confirm successful reception. After receiving the ACK, the SNs enter a deep sleep mode to conserve energy until the next UAV round.

\subsection{Data Flow Models for RI-WuR-UAC}
This subsection details the data flow model for collecting data frames from SNs within a single cluster by a single UAV. The focus is on a TSN and its interaction with the UAV, which serves as a representative communication pattern for all other SNs in the network.
 \begin{figure}[b]
	\centerline{\includegraphics[width=0.45\textwidth]{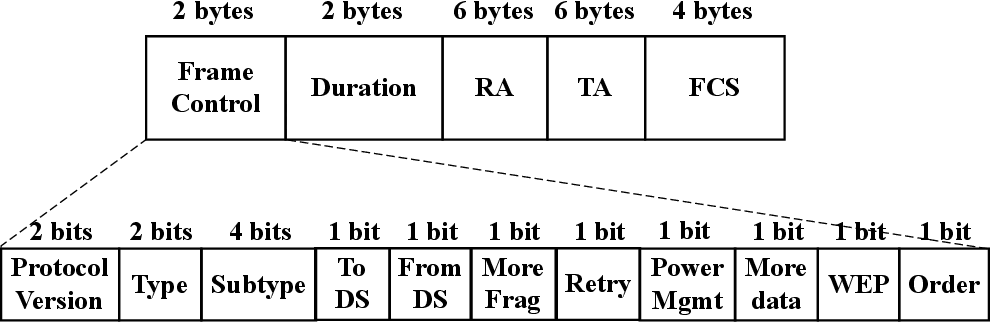}}
	\caption{Structure of JReq frame showing the assignment of bits to the frame control field.}
	\label{Ch4 JReq frame}
\end{figure}
The TSN receives a WuC signal from the UAV using its WuRx and responds by transmitting a JReq frame through its MR. The JReq frame is similar in structure to the Request to Send (RTS) frame as defined in the IEEE 802.11 standard \cite{ieee2012ieee}. It consists of five fields: frame control, duration, receiver MAC address (RA), transmitter MAC address (TA), and frame check sequence (FCS), as shown in Fig.~\ref{Ch4 JReq frame}. The figure provides a visual representation of the frame structure, with each field's size specified. The frame control field includes several subfields, and the duration field indicates the volume of data the TSN intends to transmit to the UAV. Based on this information, the UAV allocates a specific number of TDMA slots to the TSN, with each slot dedicated to collecting a single data frame from the TSN. The duration field also provides insights into the length of the JReq frame and its expected response time.

Upon receiving the JReq frame, the UAV captures the TSN's frame address and assigns TDMA slots according to the data volume specified in the duration field. The TSN then transmits its data frames within the allocated TDMA slots, and the UAV acknowledges the successful reception of these frames. Once the TSN receives the ACK from the UAV, it enters a deep sleep mode by turning off its MR, allowing it to conserve energy while still being able to sense its surroundings and listen for the next WuC signal from the UAV. The TSN discards data frames that have been successfully acknowledged due to its limited memory capacity. However, any frames that are not received or acknowledged are retransmitted in the subsequent round, ensuring reliable data transfer.
 \begin{figure*}[htbp]
	\centerline{\includegraphics[width=\textwidth]{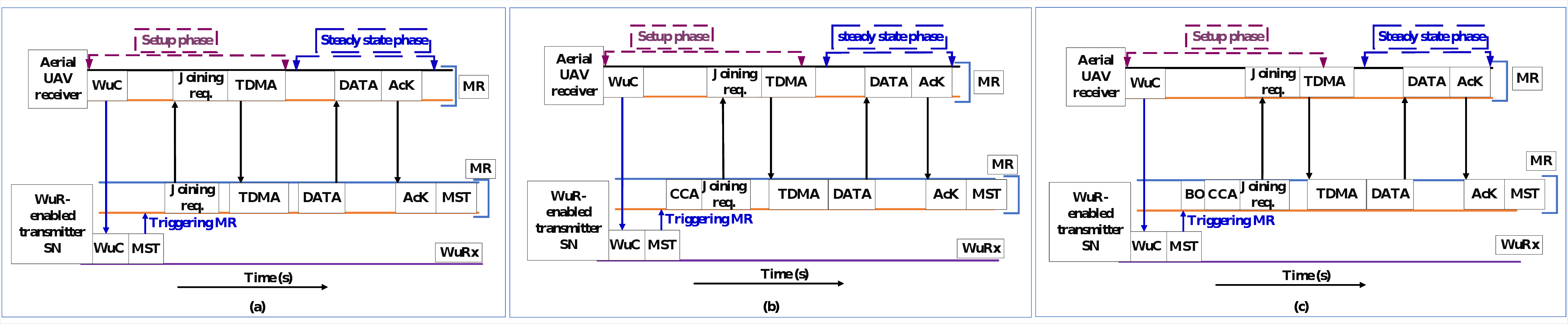}}
	\caption{Complete data flow models for the proposed RI-WuR-UAC which encloses the (a) SCM clustering with no MAC mechanism, (b) CCA clustering to support light traffic, and (c) CSMA-CA clustering to support high traffic and heavy load.}
	\label{fig2: SCM CCA & CSMA-CA Clustering RI-WuR-UAC data flow model}
\end{figure*}
In Fig. \ref{fig2: SCM CCA & CSMA-CA Clustering RI-WuR-UAC data flow model}, the complete data flow models of the proposed RI-WuR-UAC protocol are illustrated for different cluster conditions. Fig.~\ref{fig2: SCM CCA & CSMA-CA Clustering RI-WuR-UAC data flow model} (a) depicts clusters where SNs do not follow any MAC mechanism before JReq transmission, this is the case for SCM clustering. Fig.~\ref{fig2: SCM CCA & CSMA-CA Clustering RI-WuR-UAC data flow model} (b) shows the model for light clusters where SNs use CCA clustering, while Fig.~\ref{fig2: SCM CCA & CSMA-CA Clustering RI-WuR-UAC data flow model} (c) illustrates densely populated clusters utilizing CSMA-CA clustering. Each highlights the setup phase and steady-state phase for its respective scenario.

It is important to note that the data flow models assume relatively stable channel characteristics throughout a single round, as previously discussed. This assumption maximizes the likelihood of receiving ACKs for data frames from the UAV. However, if ACKs are not received, the TSN retains the data frames for retransmission in the next round. This approach minimizes the storage of additional sensed data due to limited memory capacity, thereby reducing the potential negative impacts on network performance.

\section{Modeling the RI-WuR-UAC}
\label{Sec: Protocol Modeling}
This section provides an analytical model of the proposed RI-WuR-UAC protocol using an $ M/G/1/2 $ queuing framework. The primary objective of this model is to ensure the reliable transmission of the JReq frame from the TSN to the UAV. The $ M/G/1/2 $ queuing model effectively captures key characteristics of MAC layer protocols, including a Poisson arrival process for frames from the upper network layers, a geometric distribution for frame processing time at MAC layer, a single server/channel, and a maximum queue capacity of two frames.

In the existing literature, several models have been developed to analyze the CCA and BO plus CCA algorithms. For instance, the model presented in \cite{kim2008performance} focuses on the regenerative cycle of $ M/G/1 $ queue, yet it does not account for BO periods or CCA. In contrast, our proposed model adopts $ M/G/1/2 $ queuing framework with a finite queue size for the TSN, offering a more realistic representation of network dynamics. A number of other approaches exist in the literature. For instance, the authors in \cite{malone25, 8996690} assume an infinite queue size with a constant channel busy probability post-CCA, while \cite{Pletcher26} employs a reduced buffer size using the $ M/G/1/2 $ queuing model to reflect a non-saturated scenario.

The $ M/G/1/2 $ queue is characterized by several key components. First, the arrival process describes how frames reach the head of the line (HoL), following a Poisson distribution, indicating that frames arrive independently and randomly over time. The average arrival rate, denoted by $\lambda$, remains constant throughout the analysis. Second, the processing time for frames follows a geometric distribution, which corresponds to the time required for CCA or BO and CCA, depending on the specific protocol used. This distribution represents the number of independent and identically distributed (i.i.d) CCA trials needed for JReq frame to be successfully transmitted. A successful outcome occurs when the channel is idle after CCA and the Jreq frame is transmitted or when the maximum allowed CCA attempts are reached, resulting in the JReq frame being discarded. The mean service rate of a frame arrived at MAC layer is denoted as $\mu$. Third, the $ M/G/1/2 $ queue operates with a single channel, allowing only one frame transmission at a time. When the channel is occupied, the incoming frames are queued, waiting for their turn to transmit (i.e., performing CCA or BO and CCA, which is protocol-specific). Lastly, the $ M/G/1/2 $ queue accommodates a maximum of two frames. If the channel is busy and two frames are already in the queue, any additional frames coming from the upper network layer are dropped or rejected.

To analyze the performance of the $ M/G/1/2 $ queue, several metrics are considered. First, the channel busyness probability ($\alpha$) after CCA represents the likelihood of the channel being busy after a CCA attempt, quantifying the probability that the channel is occupied and unavailable for transmission despite the CCA process. Second, the average waiting time (E[$D_{HoL}$]) measures the average time a frame spends in the queue before processing. Last is the average number of frames served E[$\tau$], quantifies the average number of frames served within a given period, considering both successful transmissions and discards, providing insight into the system's workload.

Within $ M/G/1/2 $ framework, the TSN is modeled with a regenerative busy cycle, beginning with the arrival of a frame to an empty queue and continuing until the last frame is transmitted or discarded, returning the queue to an empty state. The service time for a frame is defined as the duration from its arrival at the HoL until immediately after the last CCA attempt. The analytical model primarily focuses on modeling the constant $\alpha$, which represents the probability of the channel being sensed as busy immediately after a CCA attempt. This probability is assumed as constant throughout and is independent of the CCA attempt stage, aligning with assumptions made in previous studies \cite{ghose2019protocol, ghose2018mac, pandey2021energy, bianchi2000performance, 8996690}. This assumption allows the service time to be i.i.d.

Mathematically, $\alpha$ is defined as
\begin{equation}\label{Alpha_def}
\resizebox{\columnwidth}{!}{$
	\alpha = \frac{\textnormal{busy period of the channel (i.e., channel is served by (N-1) SNs)}}{\textnormal{total observation time (i.e., busy period + idle period)}},
 $}
\end{equation}

and the corresponding expression is
\begin{equation}\label{Alpha_equation}
	\alpha = \frac{(N-1)(1-P_{Loss})E[\tau](T_{CCA}+T_{TR})}{\frac{1}{\lambda}+E[\tau]E[D_{HoL}]}.
\end{equation}

\noindent In (\ref{Alpha_equation}), the average number of frames served in a busy cycle is represented by $E[\tau]$. The probability of discarding the JReq frame after its maximum number of CCA attempts ($MA + 1$) is denoted as $P_{Loss}$. The mean HoL delay, indicating the duration from frame arrival to the end of its last CCA attempt, is denoted as $E$[$D_{HoL}$]. Moreover, $T_{TR}$ represents the total duration of a transmission attempt, and $T_{CCA}$ signifies the time taken by a CCA attempt. The expressions for $P_{Loss}$ and $T_{TR}$ are given as

\begin{equation}\label{Prob. Loss Eq}
	P_{Loss}=\alpha ^{MA+1},
\end{equation}
and
\begin{equation}\label{trans_time}
        \resizebox{\columnwidth}{!}{$
	T_{TR}=T_{WuC}+T_{MST}+T_{JReq}+T_{TDMA}+ T_{Data} + T_{ACK}+T_{MST}.
        $}
\end{equation}
 Here, $T_\text{WuC}$ and $T_\text{TDMA}$ represent the time required for the UAV to transmit the WuC and TDMA frames, respectively. $T_\text{MST}$ and $T_\text{ACK}$ denote the time required for the TSN to switch the MR to WuRx (or vice versa) and the time to transmit the ACK frame, respectively. $T_\text{Data}$ represents the time required for the TSN to transmit $\delta$ number of frames in $n_s$ TDMA slots. The parameter $\delta$ follows a uniform random distribution, indicating the variability in the number of frames associated with the TSN. The expression for $ T_{Data} $ is given as
\begin{equation}
	T_{Data} = \delta T_T + T_G + T_{OH}.
\end{equation}
where, $ T_T $, $ T_G $, and $ T_{OH} $ represent the time required for a single data frame transmission, guard time between the two consecutive TDMA slots, and time required for overhead transmission, respectively. Their corresponding expressions are given as
\begin{equation}
	T_T = \frac{\textnormal{data frame size}}{\textnormal{data rate}},
\end{equation}

\begin{equation}
\resizebox{\columnwidth}{!}{$
	T_G = \frac{\textnormal{distance between TSN and UAV}}{\textnormal{signal propagation speed}} + \textnormal{Receiver processing time},
 $}
\end{equation}
and
\begin{equation}
	T_{OH} = \frac{\textnormal{MAC header size} + \textnormal{Security overhead}}{\textnormal{data rate}}.
\end{equation}

To completely describe $\alpha$, as given in (\ref{Alpha_equation}), the expressions for $E[\tau]$ and $E$[$D_{HoL}$] are derived below.\
 
Let $ W_j $ be the contention windows size in the $ j $th BO stage and $\sigma$ is the single slot BO duration, then the average accumulated duration until the $ j $th transmission attempt, $ w_j $, is given as
\begin{equation}\label{Eq for means acc duration}
	w_j = \sum_{v=0}^{j-1}\frac{W_v-1}{2}\sigma+jT_{CCA},
\end{equation}
where, $\sum_{v=0}^{j-1}\frac{W_v-1}{2}\sigma $ and $jT_{CCA}$ represent the mean accumulated duration of BOs and that of CCA attempts until the $ j $th transmission attempt, respectively.\

As per the definition of $ w_j $, the expression for $ E[D_{HoL}] $ can be written as
\begin{equation}\label{Eq: D-HoL}
	E[D_{HoL}]=\sum_{i=0}^{MA}\alpha^i(1-\alpha)w_{i+1}+\alpha^{MA+1}w_{MA+1}.
\end{equation}
Here, $ \sum_{i=0}^{MA}\alpha^i(1-\alpha)w_{i+1} $ and $ \alpha^{MA+1}w_{MA+1} $ represent the mean time taken by a successful transmission of JReq and discarding of JReq after ($ MA+1 $) attempts, respectively.\

Furthermore, to derive expression for average number of frames served in a busy cycle, $ E[\tau] $, it is assumed that the frame service time follows geometric distribution. Thus,

\begin{equation}\label{Prob. j packets served}
	P(\tau=j)=(1-a_0)^{j-1}a_0, \ \ \ j=1,2,3,... ,
\end{equation}
and
\begin{equation}\label{Eq Exp. No. of Packets served}
	E[\tau]=\frac{1}{a_0}
\end{equation}
where, $ a_0 $ represents the probability of the event that a frame does not encounter any other frame in the queue from its arrival at the HoL until it is successfully transmitted or discarded. The value of $ a_0 $ can be calculated as follows,
\begin{equation}\label{Eq parameter geo RV}
	a_0 \approx \sum_{i=0}^{MA}\alpha^i(1-\alpha) e ^{-(w_{i+1}+T_{TR})\lambda} +\alpha ^{MA+1}.
\end{equation}
By substituting (\ref{Eq Exp. No. of Packets served}), (\ref{Eq: D-HoL}), and (\ref{Prob. Loss Eq}) into (\ref{Alpha_equation}), a non-linear equation is obtained from which the value for $\alpha$ can be determined.

\section{Performance Matrices}
\label{Sec: Parameters for Performance Analysis}
This section derives the mathematical expressions for key performance metrics: the probability of successful cluster formation (i.e., $1 ~- ~JReq~Loss~Probability$), average SN energy consumption, and average delay. These metrics are evaluated for the three proposed MAC mechanisms: CCA clustering, CSMA-CA clustering, and ADP clustering, as well as for the benchmark SCM clustering protocol.

\subsection{JReq Loss Probability}
This parameter denotes the probability of a TSN being dropped from participating in cluster formation. When the TSN reaches the maximum allowed ($ MA +1$) failed CCA attempts, i.e., without successfully transmitting its JReq, the TSN is unable to receive a TDMA slot for its data frame transmission. Consequently, after a timeout period, the TSN enters deep sleep mode and awaits the next UAV round. This dropout probability signifies the likelihood of a TSN being excluded from the cluster in the current UAV round. In this subsection, JReq loss probability is derived for the benchmark as well as the proposed protocol as follows.

\subsubsection{SCM Clustering}
In SCM clustering, there is no BO and CCA process. When the TSN receives a WuC from the UAV, it immediately sends its JReq to the UAV without any delay. The probability of collision of the JReq frame, denoted as $\gamma$, which is evaluated by the authors in~\cite{ghose2018mac, ghose2019protocol}, is given as
\begin{equation}
	\gamma \approx 1-e^{-(N-1)\lambda T_{TR}(1+e^{(-T_{TR}\lambda)})}.
\end{equation} 
\subsubsection{CCA Clustering}
In CCA clustering, the TSN follows a simplified process for channel access. It performs CCA before transmitting JReq to the UAV, without any BO algorithm. When the TSN receives the WuC signal at its WuRx, it triggers its MCU to enable active mode. The TSN then immediately performs CCA to check if the channel is idle. If the channel is found to be idle, the TSN transmits its JReq. Otherwise, the TSN continues to perform CCA until the maximum CCA attempts ($ MA +1$) are reached. Once the $ MA +1$ is reached, the TSN discards the JReq and does not participate in the cluster formation for the current round. After a timeout period, the TSN switches to deep sleep mode and waits for the next round initiated by the UAV.\

The probability that the TSN will discard the JReq frame after MA CCA attempts can be obtained from (\ref{Prob. Loss Eq}), which can be derived using the following expression for average accumulated duration for the $ j $th transmission attempt.
\begin{equation}
	w_j = jT_{CCA}.
\end{equation}

\subsubsection{CSMA-CA Clustering}
In the case where CSMA-CA clustering is followed by the TSN for reliable JReq transmission, the JReq loss probability is given by (\ref{Prob. Loss Eq}). Before performing CCA, CSMA-CA involves a BO process. In the BO process, the TSN selects a random number of slots from a fixed CW size, where each slot has a fixed duration denoted by $\sigma$. Upon receiving the WuC from the UAV, the TSN waits according to the BO process and then performs CCA. If the channel is found to be idle after CCA, the TSN transmits its JReq. However, if the channel is found busy, the TSN follows the BO and CCA process again until the maximum CCA attempts ($ MA+1 $) are reached. Once the $ MA+1 $ is reached, the TSN discards the JReq and does not participate in the cluster formation for the current round. After a timeout period, the TSN switches to deep sleep mode and waits for the subsequent UAV round. The evaluation of $w_j$ can be expressed as
\begin{equation}
	w_j = j (\frac{CW-1}{2}\sigma+T_{CCA}).
\end{equation}
This expression is used for the calculation of $P_{Loss}$ in case where TSN follows CSMA-CA WuR.

\subsubsection{ADP Clustering}
When ADP clustering is used by the TSN to access the channel, the JReq loss probability is given by (\ref{Prob. Loss Eq}). However, ADP clustering initially follows the CCA clustering approach for the first CCA attempts. After that, it switches to the CSMA-CA clustering approach for the next three BOs followed by CCA attempts. As mentioned earlier, ADP clustering adopts CSMA-CA clustering after unsuccessful CCA attempts in CCA clustering. Therefore, the expression for $w_j$ is modified and can be represented as
\begin{equation}
        \resizebox{\columnwidth}{!}{$
	w_j = \begin{cases} jT_{CCA}, &~{\mathrm{ \forall}}~j~\in ~0,1,...,t_h-1 \\ (j+1-t_h)\frac{CW-1}{2}\sigma+kT_{CCA},&~{\mathrm{ \forall}}~j~\in ~t_h,...,MA+1. \end{cases}
 $}
\end{equation}
In this modified expression for $w_j$, $t_h$ represents the threshold number of failure attempts after which the ADP clustering switches from CCA to CSMA-CA clustering. This modified $w_j$ is considered when formulating the $P_{Loss}$ for the TSN that follows the ADP clustering protocol.

\subsection{Average Transmission Delay}
This parameter represents the average duration required by the TSN to transmit the sensed data during a single UAV round. It encompasses various stages, including the arrival of WuC, switching to active mode, performing the channel access algorithm, transmitting JReq, receiving TDMA slot frames, transmitting data frames, receiving AcK, and switching to deep sleep mode. These stages are depicted in Fig.~\ref{fig2: SCM CCA & CSMA-CA Clustering RI-WuR-UAC data flow model} for three MAC protocols. The average delay $ D_A $ is evaluated for each protocol in the following.

\subsubsection{SCM Clustering}
Since SCM clustering does not follow any BO or CCA before transmitting the TReq, the average delay is given as
\begin{equation}
	D_A= \gamma(T_{TR}-T_{ACK})+(1-\gamma)T_{TR}.
\end{equation}

\subsubsection{ CCA Clustering, CSMA-CA Clustering, and ADP Clustering}
The values for $D_A$ are different for CCA clustering, CSMA-CA clustering, and ADP clustering, as they have different values for $w_j$, which correspond to changes in $P_{Loss}$ and $E$[$D_{HoL}$]. These different values are used in the following expressions to calculate $D_A$ for CCA clustering, CSMA-CA clustering, and ADP clustering.
\begin{equation}\label{EQ: D_A}
	D_A = (1-P_{Loss})T_{tq}+P_{Loss}T_{Loss},
\end{equation}
where $ (1-P_{Loss})T_{tq} $ represents the average time taken for a successful attempt and $ P_{Loss}T_{Loss} $ represents the average time taken for a failed attempt. $ T_{tq} $ represents the average time taken by a frame from its arrival at the HoL, followed by CCA or BO and CCA, successful transmission, and reception of AcK. $ T_{Loss} $ represents the average time taken by a frame from its arrival at the HoL, performing BO and CCA, but being unable to transmit the frame. The corresponding expressions for these time durations are given, respectively, as
\begin{equation}\label{EQ: T_tq}
	T_{tq} = \frac{E[D_{HoL}]-P_{Loss}T_{Loss}}{1-P_{Loss}}+T_{TR},
\end{equation}
and 
\begin{equation}
	T_{Loss} = \sum_{i=0}^{MA}\frac{CW_i-1}{2}\sigma+(MA+1)T_{CCA}.
	\label{EQ: T_Loss}
\end{equation}
By substituting (\ref{EQ: T_tq}) and (\ref{EQ: T_Loss}) into (\ref{EQ: D_A}), the average delay or the time taken by an SN to transmit its data to the UAV can be obtained.\

As CCA-WuR does not perform any BO algorithm, hence, the expression for $ T_{Loss} $ is different from (\ref{EQ: T_Loss}) and given as
\begin{equation}
	T_{Loss} = (MA+1)T_{CCA}.
\end{equation}

\subsection{Total Energy Consumption During a Single Round}
The parameter $E_R$ represents the average amount of energy consumed by the TSN during a single UAV round. It encompasses various energy-consuming activities, including the arrival of WuC at the HoL, TSN switching to active mode, transmitting JReq, receiving the TDMA slot, transmitting data frames, receiving the AcK frame, and switching back to deep sleep mode. In this subsection, the energy consumption expressions for each MAC protocol considered are evaluated individually.

\subsubsection{SCM Clustering} 
As SCM-WuR does not perform any BO or CCA, the expression for the energy consumed by a TSN in a single round can be simplified. It is given as,
\begin{equation}
	E_{R} = \gamma (E_{TR}-E_{ACK})+(1-\gamma)E_{TR}.
\end{equation}
Here, $ E_{TR} $ represents the energy consumed by the TSN during a successful transmission attempt, and it is given by the following expression,
\begin{equation}
    \resizebox{\columnwidth}{!}{$
    E_{TR}=E_{WuC}+E_{MST}+E_{JReq}+E_{TDMA}+\delta E_{Data}+E_{ACK}+E_{MST}.
    $}
\end{equation}
Here $ E_{WuC} $, $ E_{MST} $, $ E_{JReq} $, $ E_{TDMA} $, $ \delta E_{Data} $, and $ E_{ACK} $ represent the amount of energy consumed during the reception of WuC, mode switching, transmission of JReq, reception of TDMA slot, transmission of data frames (multiplied by the number of data frames $\beta$), and reception of ACK, respectively.

\subsubsection{CCA Clustering, CSMA-CA Clustering, and ADP Clustering}
In the case of CSMA-CA and ADP clustering, the energy consumed by the TSN during BO, CCA, in each CCA stage is considered. Therefore, the expression for $ E_R $ is given as follows,
\begin{equation}\label{Eq: E_R}
	E_R = (1-P_{Loss})E_{tq}+P_{Loss}E_{Loss}.
\end{equation}
The total energy consumed by the TSN during BO, CCA, and successful data frame transmission is expressed as $ E_{tq} $, while $ E_{Loss} $ represents the energy consumed during a failed attempt. Now
\begin{equation}\label{EQ: E_tq}
	E_{tq} = \frac{E_{HoL}-P_{Loss}E_{Loss}}{1-P_{Loss}}+E_{TR},
\end{equation}
and 
\begin{equation}
	E_{Loss} = \sum_{i=0}^{MA}\frac{CW_i-1}{2}\sigma+(MA+1)E_{CCA}.
	\label{EQ: E_Loss}
\end{equation}
$ E_{HoL} $ for CSMA-CA clustering and ADP clustering is given as,
\begin{equation}
\resizebox{\columnwidth}{!}{$
	E_{HoL}  = \sum_{v=0}^{M}\alpha^v(1-\alpha)\sum_{j=0}^{M}\frac{W_j-1}{2}E_{bo} \quad +(MA+1)E_{CCA}+\alpha^{MA+1}E_{Loss}.
$}
\end{equation}

By substituting the expressions from (\ref{EQ: E_Loss}) and (\ref{EQ: E_tq}) into (\ref{EQ: D_A}), the average energy consumed by the TSN in a single UAV round can be calculated.\

Furthermore, in case where TSN follows CCA clustering for the transmission of JReq, there is no BO algorithm followed. Therefore, the following equations are respectively used for $E_{Loss}$ and $ E_{HoL} $:
\begin{equation}
	E_{Loss} = (MA+1)E_{CCA}.
\end{equation}
and
\begin{equation}
	E_{HoL}  = (MA+1)E_{CCA}+\alpha^{MA+1}E_{Loss}.
\end{equation}

\section{Simulations Results and Discussions}
\label{Sec: Simulations}
This section presents and discusses the simulation results of the three data flow models tailored for network scenarios with light traffic, heavy traffic, and varying traffic loads, each incorporating distinct channel access mechanisms: CCA clustering, CSMA-CA clustering, and ADP clustering. These are compared with the benchmark SCM clustering protocol. The simulations were conducted using MATLAB 2023b, and the results provide valuable insights into the performance of the proposed approaches. The key performance metrics considered include the probability of tagged TSN participation in cluster formation, the average time required for the TSN to transmit its sensed data, and the energy consumption of the TSN during cluster formation and data transmission.

To evaluate the performance of the protocols, the frame arrival rate at the HoL, $\lambda$, was set to $10$ frames per second. The number of SNs, $N$, in each cluster was varied from $5$ to $100$, in increments of $1$, to represent various traffic loads. This range of $N$ values is useful for analyzing the protocol's performance under different network load conditions. The remaining parameters used in the simulations are listed in Table~\ref{Tab: Ini Parameters}.\

\begin{table}[t]
	\centering 
	\caption{Initial parameters~\cite{ghose2018mac,ghose2019protocol, oller2015has, ghose2019enabling}.\label{Tab: Ini Parameters}}
	
	\begin{tabular}{|>{\centering\arraybackslash}m{1.5cm}|>{\centering\arraybackslash}m{4cm}|>{\centering\arraybackslash}m{1.5cm}|}
		\hline
		\textbf{Radio type} & \textbf{Parameter description}	& \textbf{Value}\\
		\hline \hline
		Common & Voltage	& $ 3 $ V \\
		\hline
		MR	& Data rate	& $ 250 $  kbps  \\
		\hline
		MR	& Reception current		&    $ 18.8 $	$m$A \\
		\hline
		MR	& Transmission current 	& $ 17.4 $	$ m $A \\
		\hline
		MR	& Payload size (data)	& $ 35 $ bytes \\
		\hline
		MR	& Joining request frame size	& $ 20 $ bytes \\
		\hline
		MR	& Idle current	& $ 20 $	 $\mu$A  \\
		\hline
		MR	& Size of ACK frame	& $ 11 $	  bytes \\
		\hline
		WuRx	& WuC duration	& $ 12.2 $	 $ m $s \\
		\hline
		WuRx	& CCA duration	& $ 1.92 $  $ m $s \\
		\hline
		WuRx	& Transmission current  & $ 152 $ $ m $A \\
		\hline
		WuRx	& Reception current	& $ 8 $ $\mu$A \\
            \hline
            WuRx	& Size of WuC packet	& $ 4  $	bytes	 \\
            \hline
		---	& Contention window slot duration & $ 320 $ $\mu$s \\
		\hline
		---	& Back-off current		& $ 5.16 $	 $ m $A \\
		\hline
		---	& MCU current  &  $ 2.7 $  $ m $A \\
		\hline
		---	& Mode switching time & $ 1.79 $ $ m $s   \\
		\hline
		---	& Maximum number of CCA or CCA plus BO attempts ($macMaxCSMABackoffs$) & $ 7 $		 \\
		\hline
		---	& Size of contention window 	& $ 32 $		 \\
            \hline
		---	& Minimum back-off exponent ($macMinBE$) 	& $ 3 $		 \\
		\hline
	\end{tabular}
\end{table}
\begin{figure}[!t]
	\centering
	\includegraphics[width=0.5\textwidth]{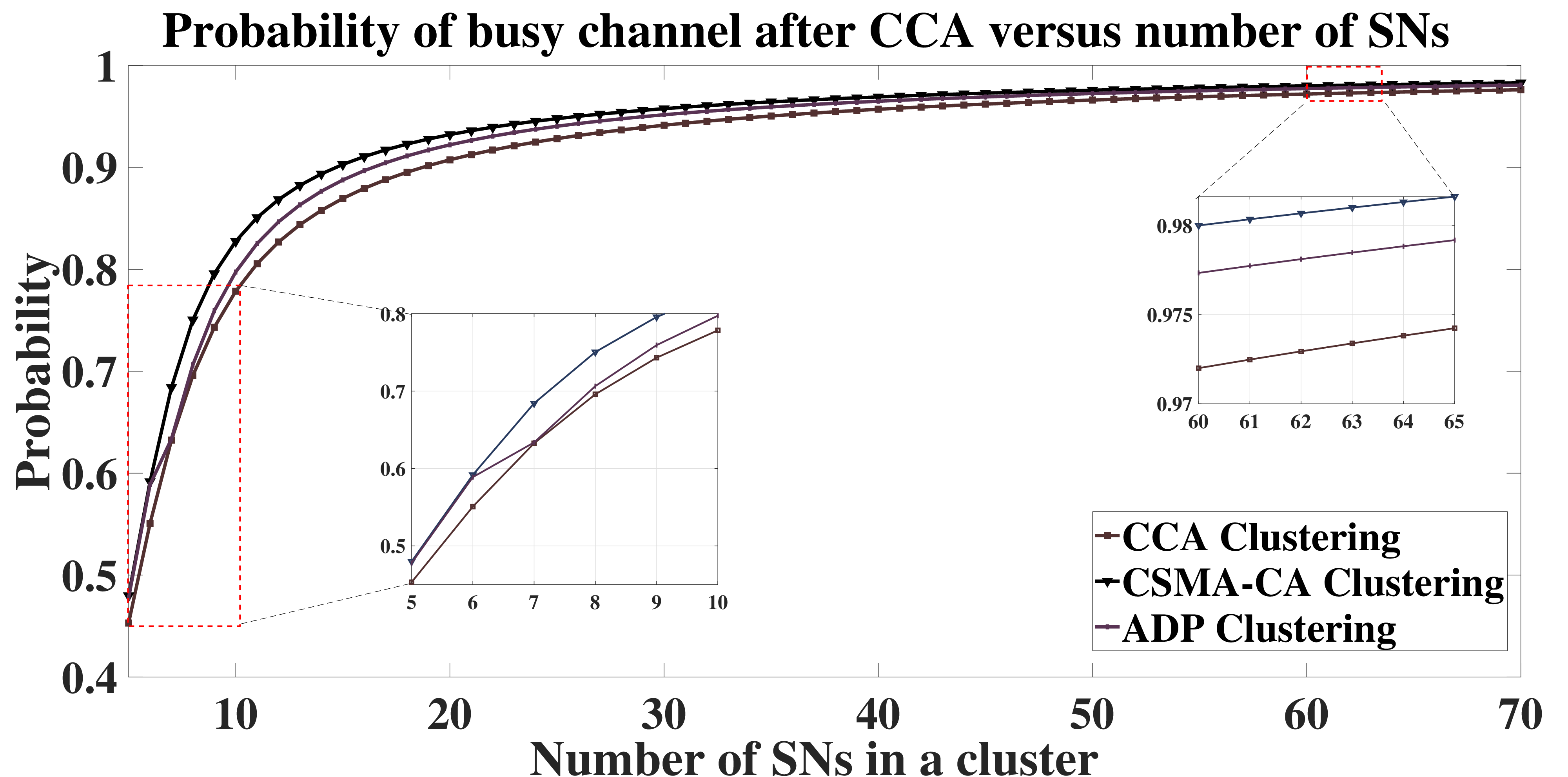}
	\caption{\centering{Channel busyness probability after performing CCA versus number of SNs.}\label{Fig: Chap 4 after CCA}}
\end{figure}

\subsection{Successful Cluster Formation}
The probability of an TSN participating in cluster formation is crucial because it determines whether the TSN will be allocated TDMA slots for reliable data transmission to the UAV. If the TSN is part of the cluster, it can transmit its sensed data frames to the UAV; otherwise, it will retain the data for transmission during the subsequent UAV round.\

Fig.~\ref{Fig: Chap 4 after CCA} illustrates the probability of the shared wireless channel being busy immediately after performing a CCA for the three proposed contention-based channel access techniques. As shown in Fig.~\ref{Fig: Chap 4 after CCA}, as the number of SNs increases, indicating a higher traffic load, the probability of channel busyness also increases. Notably, the CCA clustering protocol exhibits a lower probability of channel busyness compared to both the ADP clustering and CSMA-CA clustering protocols. This is because CCA clustering does not employ a BO algorithm. In contrast, ADP clustering initially follows the CCA clustering protocol for the first few CCA attempts but then switches to the CSMA-CA protocol, which requires a BO process before performing CCA. Furthermore, the adaptive nature of ADP clustering causes it to initially behave like CCA clustering under low network traffic conditions. However, as the traffic load increases, its behavior becomes more similar to CSMA-CA clustering.
 
\begin{figure}[t]
	\centering
	\includegraphics[width=0.5\textwidth]{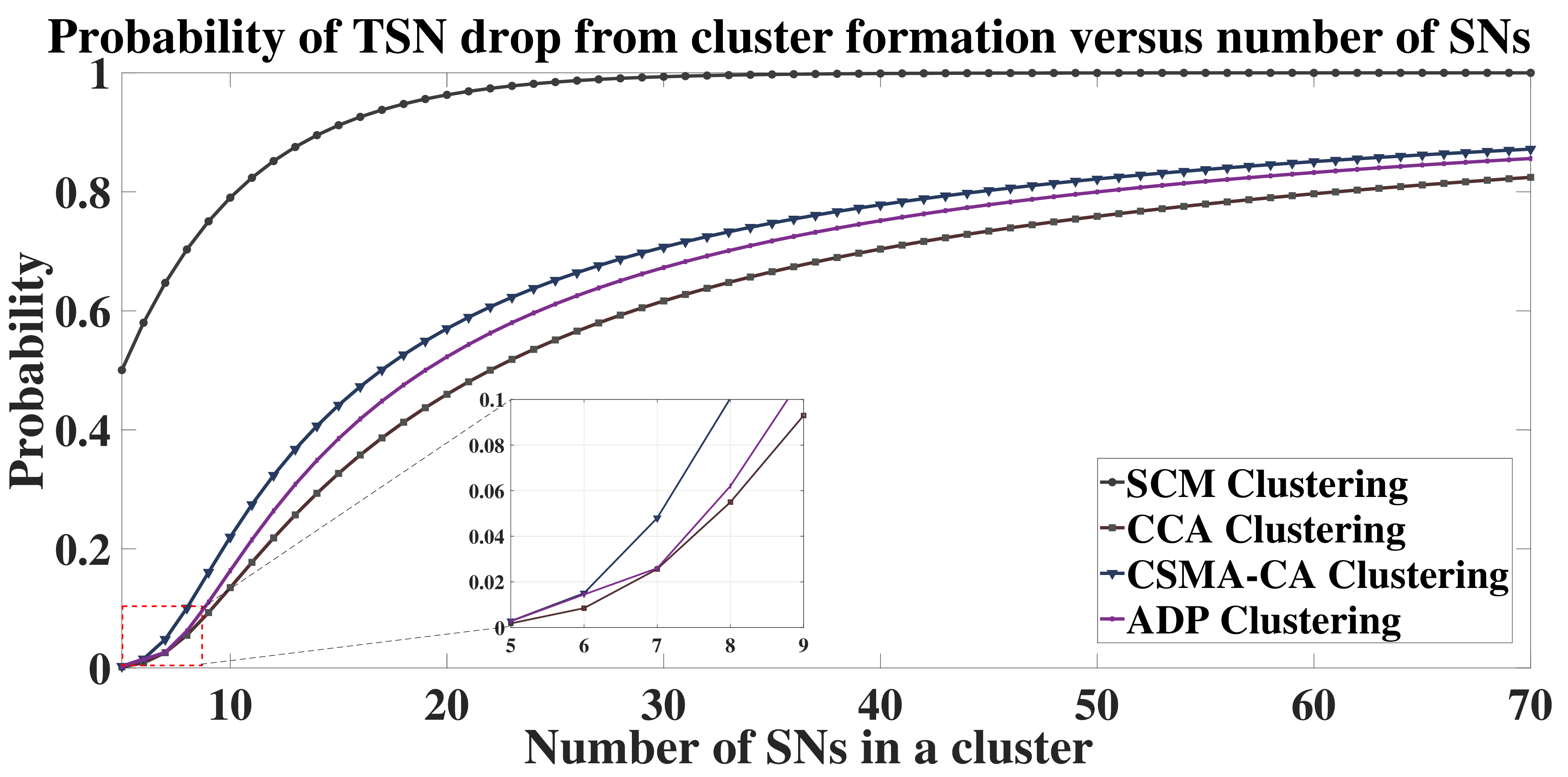}
	\caption{\centering{Probability of TSN drop from cluster formation versus number of SNs.}\label{Fig: Chap 4 SN Drop}}
\end{figure}

Fig.~\ref{Fig: Chap 4 SN Drop} shows the probability of a TSN dropping the JReq frame and failing to transmit it when the maximum number of CCA attempts is reached. This probability indicates the TSN's inability to win the channel access and transmit the JReq frame to the UAV, leading to its exclusion from participating as a CM. The figure illustrates that as the traffic load increases, the probability of JReq frame dropping also rises. Compared to the three proposed MAC protocols, the benchmark SCM clustering exhibits the highest probability of frame dropping. This is because SCM clustering does not implement any BO or CCA mechanisms, which results in lower performance for successful JReq transmission. The behavior of ADP clustering is consistent with the observations in Fig.~\ref{Fig: Chap 4 after CCA}. Initially, ADP clustering aligns with CCA under low traffic load, but as the number of SNs in the cluster increases, its behavior transitions towards that of CSMA-CA clustering.

\subsection{Average Transmission Delay}
\begin{figure}[t]
	\centering
	\includegraphics[width=0.5\textwidth]{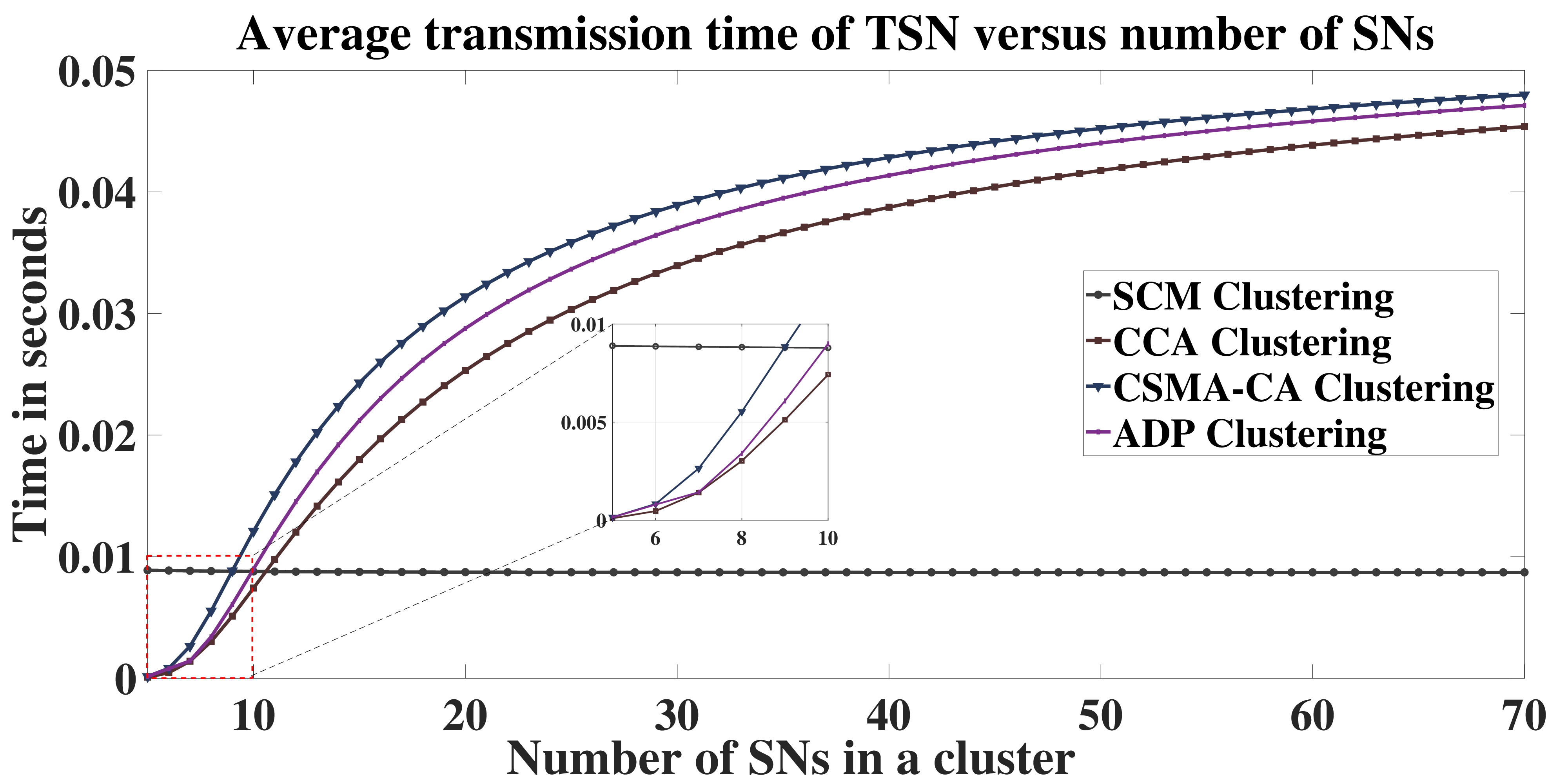}
	\caption{\centering{Average time required for TSN to transmit its sensed data versus number of SNs.}\label{Fig: Chap 4 time}}
\end{figure}

Fig.~\ref{Fig: Chap 4 time} depicts the average delay parameter, $D_A$, under varying traffic loads for the proposed three clustering MAC protocols and benchmark SCM clustering.

SNs operating under the SCM clustering protocol experience a constant average transmission delay, which remains unaffected by the traffic load. This consistancy arises because the SCM protocol does not utilize BO or CCA mechanisms, nor does it support retransmissions. The frame arrival rate, $\lambda$, is equal to the transmission or discard rate. In other words, whenever a frame arrives from the upper layer, it is transmitted if the channel is found to be free; otherwise, the frame is discarded. This process leads to a consistent transmission delay regardless of traffic conditions.

In contrast, the proposed protocols show increasing delays as the traffic load increases. CCA clustering, which solely performs CCA, offers lower delay compared to the contention-based protocols. Among the contention-based protocols, CSMA-CA clustering, which combines both BO and CCA, results in longer delays due to the additional time required for the BO process before CCA.\

ADP clustering initially follows the behavior of CCA clustering for the initial failed CCA attempts and then switches to CSMA-CA clustering. As a result, ADP clustering generally offers lower delays compared to CSMA-CA clustering but slightly higher delays compared to CCA clustering. It strikes a balance between the two protocols. The figure depicts that as the traffic load increases, the behavior of ADP clustering gradually shifts from that of CCA clustering to CSMA-CA clustering.

\subsection{Sensor Node Energy Consumption}
\begin{figure}[t]
	\centering
	\includegraphics[width=0.5\textwidth]{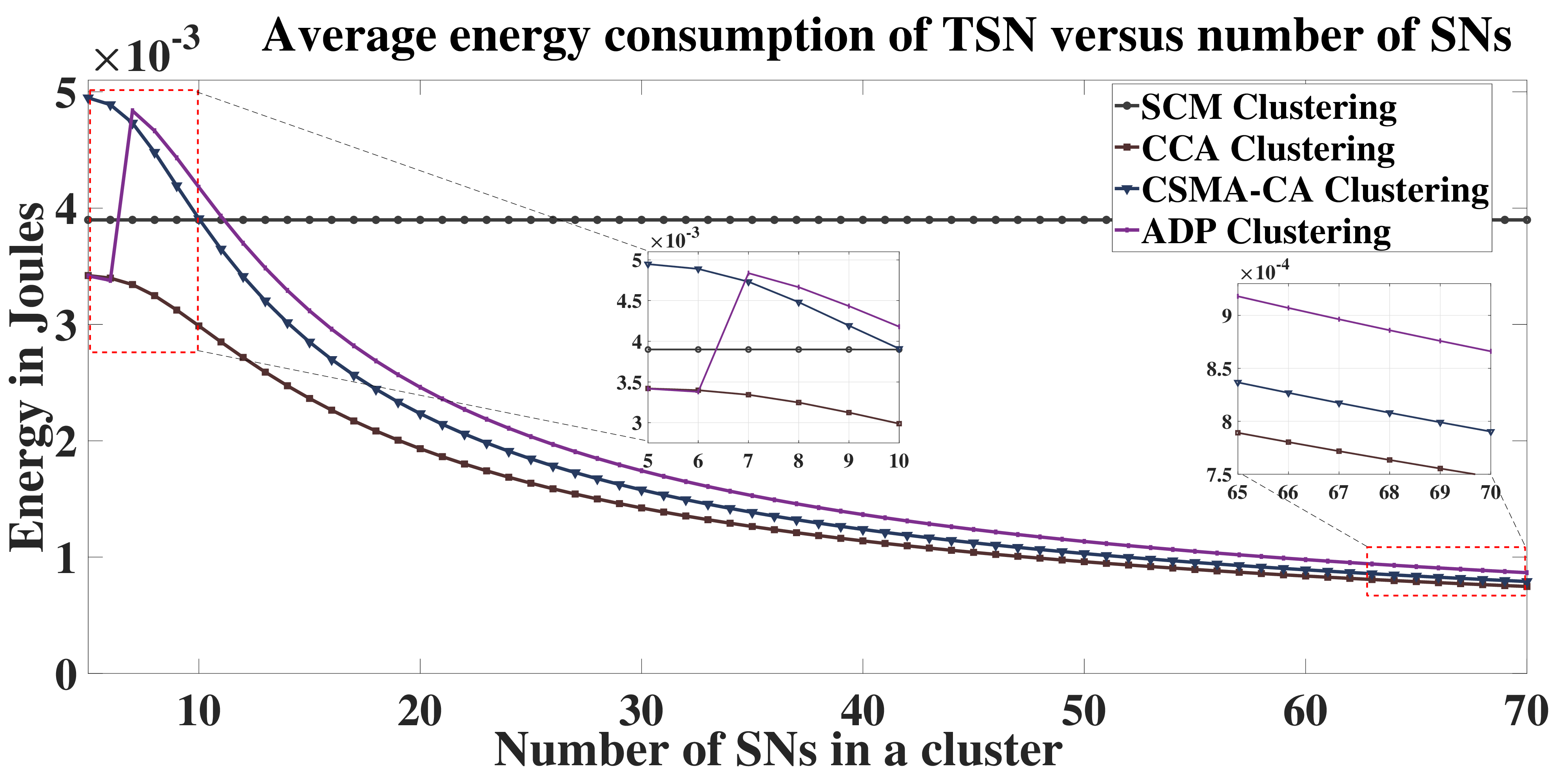}
	\caption{\centering{Average amount of energy consumed by the TSN in a single UAV round versus number of SNs.}\label{Fig: Chap 4 Energy}}
\end{figure}

Fig.~\ref{Fig: Chap 4 Energy} illustrates the average energy consumed by a TSN during a single UAV round for the proposed MAC protocols under varying traffic load conditions.

The energy consumption of TSNs following the benchmark SCM clustering protocol is both higher and constant compared to that of the CCA clustering, CSMA-CA clustering, and ADP clustering WuR MAC protocols. In SCM clustering, energy is used for both successful and failed or collided transmissions of the JReq frame. When a frame arrives at the HoL, it is immediately transmitted if the channel is free. However, if the channel is busy, a collision occurs and the transmission fails, as demonstrated in Fig.~\ref{Fig: Chap 4 SN Drop}. This results in the same transmission energy consumption for both successful and unsuccessful transmissions. Because SCM clustering does not implement BO or CCA before transmission, a significant amount of energy is wasted when frames are lost during transmission.

Conversely, the proposed contention-based clustering MAC protocols exhibit lower energy consumption compared to SCM clustering. As illustrated in Fig.~\ref{Fig: Chap 4 Energy}, the average energy consumption decreases as the number of SNs in the cluster increases, which correlates with an increase in traffic load. With more SNs competing for channel access, the probability of a TSN successfully transmitting the JReq frame decreases. Once the maximum number of CCA failure attempts is reached, the TSN fails to transmit the JReq frame, is excluded from cluster formation, and does not consume additional energy for JReq frame transmission. In such cases, the energy required for the transmission of JReq frame is greater than the energy consumed during the CCA or BO plus CCA processes. Consequently, the TSN remains stuck in the CCA or BO plus CCA process until the maximum number of CCA attempts is reached, awaiting an opportunity to win the channel access and transmit the JReq frame. This results in a linear energy consumption pattern, primarily due to the energy consumed during BO and CCA operations under heavy network conditions.

Among the contention-based clustering MAC protocols, ADP clustering consumes the most energy. Under lower load conditions, its energy consumption is similar to that of CCA clustering. However, as the load increases, ADP clustering switches to CSMA-CA clustering before reaching the initial CCA failure attempts that follow CCA clustering. The energy consumption of CSMA-CA clustering is higher than that of CCA clustering because it includes the BO process in addition to CCA.

\section{Conclusions and Potential Future Directions}
\label{Sec: Conclusions}
This paper proposed the RI-WuR-UAC protocol for UAV-assisted clustering in WuR-enabled IoT networks, designed to provide low latency, energy efficiency, and reliable data transmission in ultra-low power IoT applications. The protocol was analytically modeled using the $M/G/1/2$ queuing framework, yielding key performance metrics such as channel busyness probability, successful cluster formation probability, average energy consumption, and transmission delay.

Three distinct data flow models were introduced to handle varying traffic conditions: CCA clustering for light traffic, CSMA-CA clustering for dense traffic, and ADP clustering for variable loads. These models were evaluated against the contention-free SCM protocol. Simulation results showed that the proposed RI-WuR-UAC protocol significantly reduces SN energy consumption and improves cluster formation success rates compared to SCM, though at the cost of increased transmission delays. For example, in a network of 50 SNs, CCA clustering showed the highest energy efficiency with an average consumption of 0.9603 mJ per SN, while ADP clustering achieved the lowest transmission delay among contention-based protocols.

The RI-WuR-UAC protocol also addressed critical challenges in our earlier EEUCH routing protocol, particularly in medium access, collision avoidance, and handshaking during cluster formation. Future research will focus on optimizing the UAV-to-SN ratio for efficient data collection, ensuring reliability, low latency, and energy efficiency. A well-calibrated UAV-to-SN ratio can mitigate the increased transmission delays observed in the proposed data flow models compared to the SCM protocol. Furthermore, UAV energy consumption must be fully accounted for in order to optimize network-wide performance.

\end{document}